\documentclass[fleqn]{article}
\usepackage{espcrc2}
\usepackage{graphicx}
\usepackage[figuresright]{rotating}
\newcommand{\BR}{{\cal B}}

\newcommand{\psip}{\psi(2S)}
\newcommand{\psp}{\psi(2S)}

\newcommand{\jpsi}{J/\psi}

\newcommand{\EE}{e^+e^-}

\newcommand{\PP}{\pi^+\pi^-}
\newcommand{\KK}{K^+K^-}

\newcommand{\kskl}{K^0_SK^0_L}

\newcommand{\jpsito}{J/\psi \rightarrow }
\newcommand{\pspto}{\psi(2S) \rightarrow }

\newcommand{\M}{\frac{\sqrt{3}}{2}M}
\newcommand{\beq}{\begin{equation}}
\newcommand{\eeq}{\end{equation}}
\newcommand{\beqn}{\begin{eqnarray}}
\newcommand{\eeqn}{\end{eqnarray}}
\newcommand{\beqns}{\begin{eqnarray*}}
\newcommand{\eeqns}{\end{eqnarray*}}
\newcommand{\bfg}{\begin{figure}}
\newcommand{\efg}{\end{figure}}
\newcommand{\bitm}{\begin{itemize}}
\newcommand{\eitm}{\end{itemize}}
\newcommand{\bnum}{\begin{enumerate}}
\newcommand{\enum}{\end{enumerate}}
\newcommand{\btbl}{\begin{table}}
\newcommand{\etbl}{\end{table}}
\newcommand{\btbu}{\begin{tabular}}
\newcommand{\etbu}{\end{tabular}}

\title{Relative phase between strong and electromagnetic 
amplitudes in $\pspto 0^-0^-$ decays}
\author{C.~Z.~Yuan \address[IHEP]{Institute of High Energy Physics,
P.O.Box 918, Beijing 100039, China}
\thanks{Supported by 100 Talents Program of CAS (U-25)},
P.~Wang \addressmark[IHEP],
X.~H.~Mo\addressmark[IHEP]$^,$\address[CCAST]{China Center of Advanced
Science and Technology, Beijing 100080, China} }

\date{\today}
\begin{document}

\begin{abstract}
With the known branching ratios of $\pspto \PP$ and $\pspto \KK$, the
branching ratio of $\pspto \kskl$ is calculated as a function of the
relative phase between the strong and the electromagnetic amplitudes of the
$\psip$ decays. The study shows that the branching ratio of $\pspto \kskl$
is sensitive to the relative phase and a measurement of the $\kskl$
branching ratio will shed light on the relative phase determination in 
$\pspto 0^-0^-$ decays.
\vspace{1pc}
\end{abstract}
\maketitle

\section{Introduction}

The relative phase between the strong and the electromagnetic
amplitudes of the charmonium decays is a basic parameter in understanding
the decay dynamics. Studies have been carried out for many $\jpsi$ two-body
decay modes: $1^-0^-$~\cite{dm2exp,mk3exp}, $0^-0^-$~\cite{a00,lopez,a11},
$1^-1^-$~\cite{a11} and $N\overline{N}$~\cite{ann}. These analyses revealed
that there exists a relative orthogonal phase between the strong and the
electromagnetic amplitudes in $\jpsi$
decays~\cite{dm2exp,mk3exp,a00,lopez,a11,ann,suzuki}. 

As to $\psip$, it has been argued~\cite{suzuki} that the only large energy
scale involved in the three-gluon decay of charmonia is the charm quark
mass, one expects that the corresponding phase should not be much different
between $\jpsi$ and $\psip$ decays. There is also a theoretical argument
which favors the $\pm90^\circ$ phase~\cite{gerard}. This large phase follows
from the orthogonality of the three-gluon and one-photon virtual processes. 
But an extensively quoted work~\cite{suzuki} found that a fit to
$\psip\rightarrow 1^-0^-$ with a large phase $\pm 90^\circ$ is virtually 
impossible and concluded that the relative phase between the strong and the
electromagnetic amplitudes should be around 180 degree\footnote{In
Ref.~\cite{suzuki}, the phase $\delta=0^\circ$ between the strong amplitude and
the $negative$ electromagnetic amplitude is corresponding to the phase
$\phi=180^\circ$ between the strong amplitude and the $positive$
electromagnetic amplitude here.}.

However, it is pointed out in Ref.~\cite{wymz} that the contribution of the
continuum process via virtual photon was neglected in almost all the data
analyses in $\EE$ experiments. By including the contribution of the
continuum process, $\pspto 1^-0^-$ decays have been reanalyzed and it is
found \cite{wymphase} that the phase of $-90^\circ$ can not be ruled out.
Unfortunately the current experimental information on $\pspto 1^-0^-$ decays 
are not precise enough to determine the phase.

For the time being the experimental information for $\psip$ decays is
less abundant than that for $\jpsi$. Among the other modes used in $\jpsi$
decays to measure the relative phase, the only mode with experimental data
in $\psp$ decays is the $\pspto 0^-0^-$ ($i.e.$ pseudoscalar meson pairs),
including $\pspto \PP$ and $\pspto \KK$. 
But this is not enough to extract the phase between the strong and the
electromagnetic amplitudes, since there are three free parameters, namely,
the absolute values of the strong and the electromagnetic amplitudes, and
the relative phase between them. Another $0^-0^-$ decay channel $\pspto
\kskl$ is thus needed to determine all these three parameters. 

Although, as has been pointed out in Ref.~\cite{feldmann}, 
$\pspto 0^-0^-$ is allowed in leading-twist pQCD while 
$\pspto 1^-0^-$ is forbidden, the relative phases found in
these two modes may not necessarily be the same, it is still
interesting to test this since in $\jpsi$ decays, the phases in
these two modes are found to be rather similar.

In this letter, the existing experimental data on $\psip$ decays to $\PP$ and
$\KK$ are used as inputs to calculate the branching ratio of $\pspto\kskl$
as a function of the relative phase. Once $\BR(\pspto \kskl)$ is known, the
relative phase between the strong and the electromagnetic amplitudes in
$\psip\rightarrow0^-0^-$ decays could be determined based on the calculations
in this letter. 

\section{Theoretical framework}

In $\pspto 0^-0^-$ decays, the G-parity violating channel
$\PP$ is through electromagnetic process~(the contribution 
from the isospin-violating part of QCD is expected to be
small~\cite{chernyak2} and is neglected), $\kskl$
through SU(3) breaking strong process, and $\KK$ through both. As has been
observed in $\jpsito \kskl$~\cite{pdg}, the SU(3) breaking strong decay
amplitude is not small. Following the convention in Ref.~\cite{a11}, the
$\psp\rightarrow 0^-0^-$ decay amplitudes are parametrized as 
\beq
\begin{array}{lcl}
 A_{\PP}  &=& E~~,     \\
 A_{\KK}  &=& E + \M~~,\\
A_{\kskl} &=& \M~~, 
\end{array}
\label{em}
\eeq
where $E$ denotes the electromagnetic amplitude and $\M$ the SU(3) breaking
strong amplitude. 

As has been discussed in Refs.~\cite{wymz,wymplb}, if $\psip$ is produced in
$\EE$ experiment, the contribution of the continuum must be included in the
total amplitude, that is 
\beq
\begin{array}{lcl}
A_{\PP}^{tot}   &=& E_c + E~~,  \\
A_{\KK}^{tot}   &=& E_c + E + \M~~,   \\
A_{\kskl}^{tot} &=& \M~~,
\end{array}
\label{emc}
\eeq
where $E_c$ is the amplitude of the continuum contribution.
Besides the common part, $E_c$, $E$ and $\M$ can be expressed explicitly as
\beq
\begin{array}{lcl}
E_c &\propto&{\displaystyle \frac{1}{s}}~~,{\rule[-3.5mm]{0mm}{7mm}} \\
E   &\propto&{\displaystyle \frac{1}{s} B(s)}~~, \\
\M  &\propto&{\displaystyle {\cal C} e^{i \phi} \cdot \frac{1}{s} B(s) }~~, 
\end{array}
\label{ecem}
\eeq
where the real parameters $\phi$ and ${\cal C}$ are the relative phase and
the relative strength between the strong and the electromagnetic amplitudes,
and $B(s)$ is defined as 
\beq
B(s)=\frac{3\sqrt{s}\Gamma_{ee}/\alpha}{s-M^2_{\psip}+iM_{\psip}\Gamma_t}~~.
\label{bexpr}
\eeq
Here $\sqrt{s}$ is the center of mass energy, $\alpha$ is the QED fine
structure constant; $M_{\psip}$ and $\Gamma_t$ are the mass and the total
width of $\psip$; $\Gamma_{ee}$ is the partial width to $\EE$.

The Born order cross sections for the three channels are thus
\beqn
\lefteqn{\sigma^{Born}_{\PP}(s)= 
\frac{4\pi\alpha^2}{s^{3/2}}[1+2\Re B(s)+|B(s)|^2]} \nonumber \\
& & \times |{\cal F}_\pi (s)|^2{\cal P}_{\PP} (s)~~, \label{bnpp}
\eeqn
\beqn
\lefteqn{\sigma^{Born}_{\KK}(s)=
\frac{4\pi\alpha^2}{s^{3/2}}[1+2\Re ({\cal C}_{\phi} B(s))+|{\cal C}_{\phi}
B(s)|^2]} \nonumber \\ 
& & \times |{\cal F}_\pi (s)|^2{\cal P}_{\KK} (s)~~, \label{bnkk}
\eeqn
\beq
\sigma^{Born}_{\kskl}(s)= \frac{4\pi\alpha^2}{s^{3/2}} {\cal C}^2 |B(s)|^2
|{\cal F}_\pi (s)|^2{\cal P}_{\kskl} (s) , 
\label{bnkskl}
\eeq
where ${\cal F}_\pi (s)$ is the pion form factor and the phase space factor
${\cal P}_f (s)$ 
($f=\PP,\KK,\kskl$) is expressed as
$$
{\cal P}_{f}(s) = \frac{2}{3s}q^3_{f}~~,
$$
with $q_{f}$ the momentum of the final state particle in two-body decay.
The symbol ${\cal C}_{\phi} \equiv 1+ {\cal C} e^{i \phi} $ is 
introduced for briefness. 

For the measurement of the narrow resonance like $\jpsi$ and $\psip$ in $\EE$
experiment, the radiative correction and the energy spread of the collider
must be considered in the calculation of the observed cross sections. In
fact, the observed cross sections and the proportions of the contributions
from resonance and continuum depend sensitively on the experiment
conditions~\cite{wymplb}. 
For $\psip$ decays to $\PP$ and $\KK$, the
contributions of the continuum, as well as interference terms, must be
subtracted from the total cross sections to obtain the correct branching
ratios. For $\kskl$ mode, there is no continuum contribution. Although 
the observed $\kskl$ cross section depends on the energy spread, the
branching ratio is simply the observed $\kskl$ cross section divided by the
total resonance cross section. The formulae to calculate the experimentally
observed cross section are presented in Ref.~\cite{wymplb}. In the following
analysis, the energy spread of different $\EE$ colliders, as listed in
Table~\ref{delta3}, are adopted in the corresponding calculations. In
addition, it is also assumed that experimental data are taken at the energy
which yields the maximum inclusive hadronic cross section~\cite{wymplb}. 

\begin{table}
\caption{\label{delta3} Energy spreads of different experiments.}
\begin{tabular}{r|ccc} \hline
  Exp.   & DASP/       & BESI/       & MARKIII/    \\
         & DORIS       & BEPC        & SPEAR       \\ \hline
$E_{cm}$ &  $\psip$    &  $\psip$    &  $\jpsi$    \\
 (GeV)   &  (3.686)    &  (3.686)    &  (3.096)    \\ \hline
 Energy  &  2.0 MeV    &  1.3 MeV    &  2.4 MeV    \\
 spread  &             &             &             \\ \hline 
\end{tabular}
\end{table}

\section{Experimental data and predictions of $\BR(\pspto \kskl)$}

Presently the experimental data on $\pspto 0^-0^-$ are limited. The
only results which have been published are from DASP~\cite{dasp}:
\beq
\BR(\pspto \PP) = (8\pm 5)\times 10^{-5}~~,
\label{dasp_pp}
\eeq
\beq
\BR(\pspto \KK) = (10\pm 7)\times 10^{-5}~~,
\label{dasp_kk}
\eeq
which are based on about $0.9\times 10^{6}$ produced $\psp$ events. 
The uncertainties of the measurements are more than 60\% because of the
small data sample. 

Another attempt to measure the branching ratios of $\pspto \PP$ and $\KK$
is based on $2.3\times 10^{6}$ $\psp$ data collected by BESI, the results
are~\cite{yesw}: 
\beq
\BR(\pspto \PP) = (0.84\pm 0.55^{+0.16}_{-0.35})\times 10^{-5},
\label{bes_pp}
\eeq
\beq
\BR(\pspto \KK) = (6.1\pm 1.4^{+1.5}_{-1.3})\times 10^{-5}.
\label{bes_kk}
\eeq
Here the uncertainty for $\PP$ is also considerably large, around 70\%; 
while for $\KK$, the uncertainty is about 30\%. 

It should be emphasized that the aforementioned values without subtracting
the contributions from the continuum are not the real physical branching
ratios. These values should be multiplied by the experimentally measured
total resonance cross section of the corresponding experiment and the
products are to be interpreted as the observed cross sections of these two
modes under the particular experimental condition. More detailed discussion
of this point is in Ref.~\cite{wymplb}. 

Since in both of these two experiments, the $\PP$ branching ratios have
large uncertainties, and the central values differ by almost an order of
magnitude, an alternative way to do the analysis is to estimate 
$\BR(\psip\rightarrow\PP)$ in terms of pion form factor extrapolated from
$\BR(\jpsi\rightarrow\PP)$ with better precision. For this purpose,
$\BR(\jpsi\rightarrow\PP)$ = $(1.58\pm 0.20\pm 0.15) \times 10^{-4} $ from
MARKIII/SPEAR~\cite{mk3pp} is used. Although the contribution of the
continuum is small for $\jpsi$ decays, it is taken into account in the
calculation here which yields 
\begin{equation}
 |F_{\pi}(M^2_{\jpsi})| = (9.3\pm 0.7)\times 10^{-2}~.
\label{Fpjpsi}
\end{equation}
Extrapolate the result by $1/s$ dependence~\cite{Manohar,chernyak}
the pion form factor becomes
\begin{equation}
 |F_{\pi}(s)| = \frac{(0.89\pm0.07)\mbox{~GeV$^2$}}{s}~~.
\label{ff_pp}
\end{equation}

With the pion form factor in Eq.~(\ref{ff_pp}), for example, BESI should
observe a $\PP$ cross section of 11.6~pb at $\psip$ energy, of which 4.8~pb
is from the resonance decays (the total $\psip$ cross section is 640~nb). 

With the input of the branching ratios of $\PP$ and $\KK$, the branching
ratio of $\kskl$ is calculated as a function of the phase between $E$ and
$\M$, as solved  from Eqs.~(\ref{bnpp}), (\ref{bnkk}) and (\ref{bnkskl})
with radiative correction and energy spread of the $\EE$ collider considered.
Three sets of inputs are used for the calculations : 
\begin{itemize}
 \item {\bf Input~1}: DASP results in Eqs.~(\ref{dasp_pp})
and (\ref{dasp_kk});
 \item {\bf Input~2}: BESI results in Eqs.~(\ref{bes_pp})
and (\ref{bes_kk});
 \item {\bf Input~3}: pion form factor from Eq.~(\ref{ff_pp})
and $\BR(\pspto \KK)$ from BESI measurement in Eq.~(\ref{bes_kk}). 
\end{itemize}

\begin{figure}[htb]
\centerline{\includegraphics[width=8.0cm,height=7.0cm]{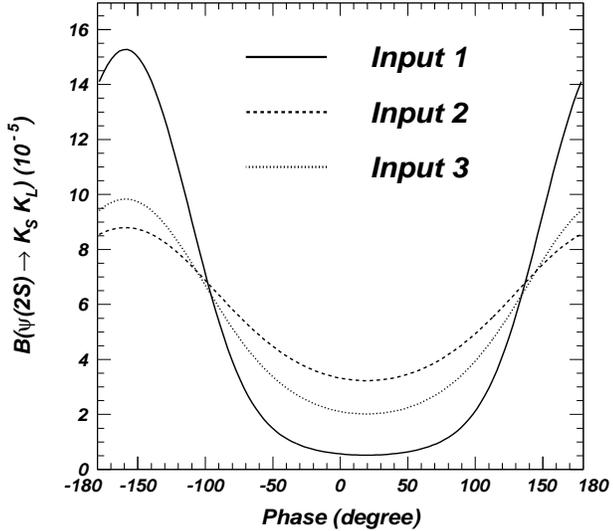}}
\caption{\label{three} $\pspto \kskl$ branching ratio as a function of the
relative phase for three different inputs which are described in the text.} 
\end{figure}

Fig.~\ref{three} shows $\BR(\pspto \kskl)$ as a function of the phase
for the three sets of inputs. It could be seen that  $\BR(\pspto \kskl)$ is
very sensitive to the relative phase. With all three sets of inputs, the
variation shows the same trend. They reach the maxima and minima at roughly
the same values of the phase. With the Input~1, $\BR (\psip\rightarrow\kskl)$
varies in a larger range than the other two sets of inputs. This is because
the $\PP$ branching ratio from DASP is large, so the electromagnetic
amplitude $E$ and the continuum amplitude $E_c$ are relatively large 
compared with the strong decay amplitude $\M$, so the interference is more
important. On the contrary, with the Input~2, the $\PP$ branching ratio is
small from BESI experiment, which means that $E$ and $E_c$ are relatively
small, so the interference is less significant. 

Table~\ref{ksklbr} lists the predictions of the $\pspto \kskl$ branching
ratios, as well as the relative strength ${\cal C}$,
with some values of the phase which are most interesting from 
theoretical point of view. These phases are $\phi=-90^{\circ}$,
$+90^{\circ}$, $180^{\circ}$ and $0^{\circ}$, for the three sets of inputs
as discussed above. The first two phases are favored by the
theory~\cite{gerard}, and are the fitted results from $\jpsi$ data; while
the third one is from an early fitting of $\psip\rightarrow1^-0^-$
mode~\cite{a00}. Here the uncertainties due to the experimental errors of
$\PP$ and $\KK$ measurements are included in the table. With the third set
of input, the theoretical uncertainty due to the extrapolation of the pion
form factor from $\jpsi$ to $\psip$ according to $1/s$ dependence is not
included. 

\begin{table}[t]
\caption{\label{ksklbr} Predicated $\BR(\pspto \kskl)$ ($\times 10^{-5}$) and 
relative strength parameter ${\cal C}=|\M/E|$ at different phases for 
different inputs.} 
\center
\begin{tabular}{r|c|ccc} \hline \hline
    Phase                                &
  &  Input~1           &  Input~2           &  Input~3   \\ \hline \hline
$-90^{\circ}$ {\rule[-3.5mm]{0mm}{9mm}}  & ${\cal B}$
  &$5.2^{+9.4}_{-5.2}$ &$6.3^{+2.2}_{-2.1}$ &$5.8^{+2.3}_{-2.2}$ \\
\mbox{}       {\rule[-3.5mm]{0mm}{7mm}}  & ${\cal C}$
  &$1.5^{+1.2}_{-1.5}$ &$4.5^{+5.1}_{-1.4}$ &$2.9^{+0.7}_{-0.6}$ \\
$+90^{\circ}$ {\rule[-3.5mm]{0mm}{9mm}}  & ${\cal B}$  
  &$1.5^{+6.9}_{-1.5}$ &$4.5^{+2.1}_{-1.9}$ &$3.4^{+1.8}_{-1.6}$ \\
\mbox{}       {\rule[-3.5mm]{0mm}{7mm}}  & ${\cal C}$
  &$0.79^{+1.94}_{-0.79}$ 
                       &$3.8^{+5.1}_{-1.4}$ &$2.2^{+0.7}_{-0.6}$ \\
$180^{\circ}$ {\rule[-3.5mm]{0mm}{9mm}}  & ${\cal B}$
  &$14^{+11}_{-14}$    &$8.6^{+2.5}_{-2.7}$ &$9.4^{+2.7}_{-2.7}$ \\
\mbox{}       {\rule[-3.5mm]{0mm}{7mm}}  & ${\cal C}$
  &$0.48^{+1.82}_{-0.48}$ 
                       &$3.3^{+5.0}_{-1.4}$ &$1.8^{+0.6}_{-0.7}$ \\
$  0^{\circ}$ {\rule[-3.5mm]{0mm}{9mm}}  & ${\cal B}$ 
  &$0.6^{+4.5}_{-0.6}$ &$3.3^{+2.2}_{-1.7}$ &$2.1^{+1.4}_{-1.2}$ \\ 
\mbox{}       {\rule[-3.5mm]{0mm}{7mm}}  & ${\cal C}$
  &$2.5^{+1.7}_{-2.5}$ &$5.2^{+5.0}_{-1.3}$ &$3.7^{+0.6}_{-0.5}$ \\ \hline \hline
\end{tabular}
\end{table}

In principle, the electromagnetic amplitudes of $\pspto \PP$ ($E_\pi$)
and $\pspto \KK$ ($E_K$) are not necessarily the same as assumed
in Eq.~(\ref{em}), a variation of $E_K$ by $\pm (20\sim 30\%)$ from $E_\pi$ 
is tested for various input. The changes of the predicted branching
ratios of $\pspto \kskl$ are well within the quoted errors since the
uncertainties of the $\BR(\pspto \PP)$ are large for Input~1 and 
Input~2; while for Input~3, the resulting branching ratio curve lies 
between the two curves from Input~1 and Input~2 in Fig.~\ref{three}.

\section{Discussions}

From Fig.~\ref{three} and Table~\ref{ksklbr}, it can be seen that
with the Input 1, the central value of $\pspto \kskl$ changes dramatically
as the phase varies. Nevertheless, such predictions 
come with huge uncertainties due to the large experimental errors of the 
input  $\BR (\pspto \PP)$ and $\BR (\pspto \KK)$. As a matter of fact,
the results by DASP in Eqs.~(\ref{dasp_pp}) and (\ref{dasp_kk}) can
accommodate the assumption within one standard  deviation that $\M=0$ in 
Eq.~(\ref{emc}), $i.e.$ the strong interaction is totally absent which
means $\BR (\pspto \kskl)=0$. Such huge uncertainties  make it 
virtually impossible to draw any useful conclusion about the phase 
even with $\BR (\pspto \kskl)$ measured.

However, with Input 2, because of the smaller error of $\BR(\pspto\KK)$ 
and the relatively small $\pspto\PP$ branching ratio, 
$\BR(\pspto \kskl)$ are calculated with much smaller uncertainty.
The strong interaction amplitude $\M$ is nonzero within two standard 
deviation, and $\BR(\pspto \kskl)$ is predicted at the order of 
$10^{-5}$. The exact value depends on the phase and varies 
by a factor 2.7 from the minimum to maximum. The uncertainty of the 
prediction, depending on the phase, is between 33\% to 50\%. 
So with this result, once $\BR (\pspto \kskl)$ is measured,
the phase between the strong and the electromagnetic amplitudes
can be determined to be within one of the following regions:
close to $0^\circ$, around $\pm90^\circ$, or close to $180^\circ$.  

With Input 3, the usage of the better measured pion form factor
at $\jpsi$ does not reduce the uncertainty 
of the predicted $\BR (\pspto \kskl)$ very much. This is due to the larger
pion form factor and so larger contribution from the electromagnetic
interactions ($E$ and $E_c$ in Eq.~(\ref{emc})) than with Input 2. 
But the predicted central values of $\BR (\pspto \kskl)$ vary in a larger
range, with a factor of 4.9 from the minimum to maximum. This makes it more
sensitive to determine the phase by $\BR (\pspto \kskl)$ than with Input~2.

By virtue of the calculations with Input~2 and Input~3, once
$\BR(\pspto \kskl)$ is known, at least it can distinguish 
whether the strong and the electromagnetic amplitudes are roughly orthogonal
(with phase around $\pm90^\circ$) or of the same or opposite phase
($0^\circ$ or $180^\circ$). This is highly desirable from the theoretical
point of view. 

To determine the relative phase between the strong and the electromagnetic
interactions with small error, the branching ratios of $\psip\rightarrow\PP$ 
and $\psip\rightarrow\KK$ must also be measured to high precisions. These
are expected from the forthcoming CLEOc and BESIII
experiments~\cite{cleoc,bes3}. 

\section{Summary}

$\pspto \kskl$ branching ratio is calculated as a function of the 
relative phase between the strong and the electromagnetic amplitudes,
based on the available experimental information of $\pspto \PP$ and
$\pspto \KK$ decay branching ratios. 
With the results in this letter, a measurement of the $\pspto \kskl$
branching ratio will shed light on answering the question that whether the
phase between the strong and the electromagnetic amplitudes is large
($\pm90^\circ$) or small ($0^\circ$ or $180^\circ$) in the 
$\psip \rightarrow 0^-0^-$ decays.

\end{document}